\documentclass[11pt,a4paper,final]{article}
\usepackage{jheppub}
\usepackage[latin1]{inputenc}

\usepackage{bbm,bm}
\usepackage{graphicx}
\usepackage{bm}
\usepackage{wrapfig}
\usepackage{amssymb}
\usepackage{mathrsfs}
\usepackage{amsmath}
\usepackage{comment}
\usepackage{amsbsy}
\usepackage{mathrsfs}
\usepackage{amsfonts}
\usepackage{epsf,color,colordvi,pifont}
\usepackage{lineno}
\usepackage{nicefrac}
\usepackage[notref,notcite]{showkeys} 

\DeclareFontFamily{OT1}{pzc}{}
\DeclareFontShape{OT1}{pzc}{m}{it}{<-> s * [1.10] pzcmi7t}{}
\DeclareMathAlphabet{\mathpzc}{OT1}{pzc}{m}{it}

\title{Axion-induced birefringence effects in laser  driven  nonlinear vacuum interaction}

\author[a,1]{Selym  Villalba-Ch\'avez %
\note{Present address: Institut f\"{u}r Theoretische Physik I, Heinrich Heine Universit\"{a}t D\"{u}sseldorf,  Universit\"{a}tsstr. 1, 40225 D\"{u}sseldorf, Germany.}}

\author[a]{and Antonino Di Piazza}

\affiliation[a]{Max-Planck-Institut f\"ur Kernphysik, Saupfercheckweg 1 D-69117 Heidelberg, Germany.}

\emailAdd{selym@tp1.uni-duesseldorf.de}
\emailAdd{dipiazza@mpi-hd.mpg.de}

\abstract{The propagation of a probe electromagnetic field through a counterpropagating strong plane wave is investigated. The effects of the electromagnetic field-(pseudo)scalar 
axion field interaction and of the self-interaction of the electromagnetic field mediated by virtual electron-positron pairs in the effective Lagrangian approach are included. First, 
we show that if the strong field is circularly polarized, contrary to the leading-order nonlinear QED effects, the axion-photon interaction induces a chiral-like birefringence and 
a dichroism in the vacuum. The latter effect is explained by evoking the conservation of the total angular momentum along the common propagation direction of probe and the strong 
wave, which allows for real axion production only for probe and strong fields with the same helicity. Moreover, in the case of ultra-short strong pulses, it is shown that the absorption 
coefficients of probe photons depend on the form of the pulse and, in particular, on the carrier-envelope phase of the strong beam. The present results can be exploited experimentally 
to isolate nonlinear vacuum effects stemming from light-axion interaction, especially at upcoming ultra-strong laser facilities, where stringent constraints on the axion-photon 
coupling constant are in principle provided.}

\arxivnumber{1307.7935}
\begin{document}
\maketitle
\flushbottom

\section{Introduction}

The ATLAS and the CMS collaboration have provided compelling evidences which point out the existence of a new heavy resonance \cite{ATLAS:2012ad,ATLAS:2012ac,Chatrchyan:2012twa,Chatrchyan:2012dg} 
that resembles  closely  the Higgs boson of the standard model \cite{Ellis:2013lra}. At energies much
below the scale specified by its mass $\approx 125\ \rm GeV$, other scalar and pseudoscalar representations of the Lorentz group are likely to occur as elementary particles, too. In some cases
they arise as an outcome of theories necessary for solving specific problems in the Standard Model. The QCD axion, for instance, is the Nambu-Goldstone boson resulting from the spontaneous
breaking of the Peccei-Quinn symmetry. It enforces the strong CP conservation \cite{Peccei:1977hh,Wilczek:1977pj,Weinberg:1977ma} and constitutes the genuine paradigm of axion-like particles
(ALPs). These are very light and weakly interacting (pseudo)-scalar particles, whose theoretical conception appears connected to some extensions of the Standard Model resulting from string 
compactifications   \cite{Witten:1984dg,Svrcek:2006yi}. Although neither axion nor ALPs have been detected so far experimentally, astrophysical and cosmological arguments have allowed 
to impose severe constraints on their masses and coupling constants. While the QCD axion might have a mass embedded between $1\;\mu\rm eV-10\;meV$ \cite{Asztalos:2001jk,lamoreaux}, the ALPs 
masses are much less constrained \cite{Jaeckel:2010ni,Hewett:2012ns}. Furthermore, the solar monitoring of a plausible axion flux from the sun \cite{CAST} as well as an assessment of its role 
in the cooling of stars \cite{Raffelt:1985nk,Raffelt:1999tx,Raffelt:2006cw} are setting the most stringent upper bound on the strength of the ALP-photon coupling $g\lesssim10^{-10}\ \rm GeV^{-1}$. 
Despite that, the validity of these limits must be considered with care   due to their dependence on the models from which they are obtained \cite{evading,Gies:2007ua}. This fact represents a strong motivation 
for investigating alternative laboratory-based routes, ideally, with enough sensitivity as to compete with the astrophysical constraints.

In order to achieve this goal, many theoretical  researches have been carried out. Most of them rely on the hypothetical photon-ALP oscillations driven by an external magnetic field 
\cite{maiani,Raffelt:1987im,Gabrielli:2006im,Gies:2007ua,Biggio:2006im}. This is because, on the one hand, the resulting axion effects manifest in the modification of the optical properties of
the vacuum, i.e., birefringence and dichroism \cite{Affleck:1987gf,Dittrich,adler,shabad4,Shabad:2011hf,VillalbaChavez:2012ea}, whereas, on the other hand, the regeneration process of a 
photon from an ALP could also provide some trace of its existence \cite{VanBibber:1987rq,Adler:2008gk,Arias:2010bh,Redondo:2010dp}. Both ideas have been implemented in polarimetry 
\cite{Cameron:1993mr,Zavattini:2007ee,BMVreport,Chen:2006cd} and ``Light Shining Through a Wall''  \cite{Ehret:2010mh,Ehret:2009sq,Chou:2007zzc,Steffen:2009sc,Afanasev:2008jt,Afanasev:2006cv,Pugnat:2007nu,Robilliard:2007bq,Fouche:2008jk}
experiments, but the  feeble  coupling between a photon and an ALP remains a big challenge to overcome  in the search of high levels of sensitivity.  There are, however,  hopes to solve 
this issue by increasing the magnetic field strength and its spatial extension. Currently, the interaction region can be macroscopically extended up to an effective
distance of the order of kilometers. But, attainable strengths of the magnetic field are of the order of $10^4\text{-}10^5\;\text{G}$, which is not large enough to make the desirable effects
manifest. With the perspectives of achieving soon much stronger field strengths of the order of $10^{11}\text{-}10^{12}\;\text{G}$ in a short linear space-extension of the order of
$10\;\text{$\mu$m}$ \cite{ELI,xcels}, ultrahigh intense lasers can become an alternative tool for investigating ALPs-induced effects \cite{mendonza,Gies:2008wv,Dobrich:2010hi,Dobrich:2010ie}.
These facilities are envisaged to reach the power level of $200\;\text{PW}$ (future prospects  even aim to reach the $1\;\text{EW}$ threshold \cite{xcels}), corresponding to a peak intensity
of the order of $10^{25}\;\text{W/cm$^2$}$ at diffraction limit. Since such intensities are only few orders of magnitudes below the critical one of QED ($I_{cr}=4.6\times 10^{29}\;\text{W/cm$^2$}$),
there are also hopes that nonlinear effects including the production of electron-positron pairs from the vacuum \cite{Hebenstreit:2009km,ruf2009zz,mocken2010}, photon splitting \cite{DiPiazza:2007yx} and
vacuum-polarization effects \cite{DiPiazza:2006pr,BenKing:2009,BenKing:2010,Heinzl1,Heinzl} may soon be detectable for the first time, together with other processes beyond QED as pion \cite{dadi} and
Higgs \cite{sarah} production (see also the recent review \cite{Di_Piazza_2012}).
However, the presence of an ALP would distort and mix with all these other effects, in particular, with those associated to the dispersive phenomena. Hence, it is desirable to find a setup where
vacuum-polarization effects as predicted by QED and mediated by virtual electron-positron pairs, and those elicited by the presence of ALPs could be ideally isolated from each other.

In this work we investigate self-interaction effects of the electromagnetic field mediated by both virtual electron-positron pairs and the ALPs. We put forward a very simple setup, which allows
in principle to isolate polarization-dependent effects stemming only from the self-interaction of the electromagnetic field in vacuum mediated by a scalar or pseudoscalar ALP. In fact, we find that if a
probe plane-wave field counterpropagates with respect to a strong, circularly polarized plane wave, the ALP-mediated interaction between the probe and the strong field depends on the mutual
helicity of the fields, whereas the one resulting at leading order \cite{Mitter,baier,mini} from the polarization of the virtual electron-positron pairs does not distinguish between the two
helicity states of the probe. In particular, we will see that in the investigated regime, due to the conservation of the total angular momentum along the common propagation direction of the fields, a real ALP can be created only if the
probe and the strong field has the same helicity.  Moreover, the ALP-photon coupling is shown to induce modifications in the  birefringence and dichroism of the vacuum of virtual electron-positron
pairs. In addition, we indicate that in the case of an ultra-short strong plane wave, ALP-induced birefringence and dichroic effects also depend on the carrier-envelope phase (CEP) of the  strong field. 
The effects calculated here could provide strong constraints on the axion-photon coupling constant if tested at upcoming ultra-intense laser facilities as the Extreme Light 
Infrastructure (ELI) and Exawatt Center for Extreme Light Studies (XCELS) but also by means of relatively long laser pulses  of more  moderate intensities.

\section{General considerations \label{GC}}

The interaction of an  axion field $\Phi(x)$  and an electromagnetic field $F^{\mu\nu}(x)=(\bm{E}(x),\bm{B}(x))$ including
nonlinear QED effects is described by the Lagrangian density\footnote{Natural and Gaussian units with $\hbar=c=4\pi\epsilon_0=1$ are employed throughout.}
\begin{equation}
\label{L}
\begin{split}
\mathscr{L}=&-\frac{1}{4\pi}\mathscr{F}+\frac{\alpha}{360\pi^2}\frac{4\mathscr{F}^2+7\mathscr{G}^2}{F_{cr}^2}+\frac{1}{2}[(\partial_{\mu}\Phi)^2-m^2\Phi^2]-\frac{g}{4\pi}\Phi(\delta_{a,s}\mathscr{F}+\delta_{a,p}\mathscr{G}).
\end{split}
\end{equation}
In this equation we have introduced: 1) the fine-structure constant $\alpha=e^2\approx 1/137$, with $e<0$ being the electron charge; 2) the two electromagnetic
invariants $\mathscr{F}(x)=F^{\mu\nu}(x)F_{\mu\nu}(x)/4=-(E^2(x)-B^2(x))/2$ and $\mathscr{G}(x)=\tilde{F}^{\mu\nu}(x)F_{\mu\nu}(x)/4=-\bm{E}(x)\cdot\bm{B}(x)$,
where $\tilde{F}^{\mu\nu}(x)=\epsilon^{\mu\nu\alpha\beta}F_{\alpha\beta}(x)/2$ is the dual tensor of $F^{\mu\nu}(x)$, with $\epsilon^{\mu\nu\alpha\beta}$
being the completely antisymmetric four-rank tensor ($\epsilon^{0123}=+1$); 3) the critical electromagnetic field of QED $F_{cr}=m_e^2/|e|$, with $m_e$ being the
electron mass; 4) the ALP mass and coupling constant $m$ and $g$, respectively; 5) the discrete variable $a$, which can be equal to either $s$ for a scalar ALP
or $p$ for a pseudoscalar ALP.

The terms in eq.~(\ref{L}) proportional to the fine-structure constant are the lowest-order terms of the Euler-Heisenberg Lagrangian
density in the ratios $|\mathscr{F}(x)|/F_{cr}^2$ and $|\mathscr{G}(x)|/F_{cr}^2$, which are assumed here much smaller than unity (we recall that the critical
field of QED corresponds to a laser peak intensity of $I_{cr}=F_{cr}^2/4\pi=4.6\times 10^{29}\;\text{W/cm$^2$}$, which exceeds by about seven orders of magnitude
presently available laser intensities \cite{Yanovsky_2008}). These terms stem from the self interaction of the electromagnetic field mediated by virtual electron-positron pairs and,
strictly speaking, arise only in the case of constant (uniform) background fields, at least at a typical time (space) scale $\lambda_C=1/m_e=1.3\times 10^{-21}\;\text{s}$ ($\lambda_C=1/m_e=3.9\times 10^{-11}\;\text{cm}$) \cite{Mitter,baier,mini}. We assume that the electromagnetic field
here contains only frequencies (wavelengths) such that the above assumption is justified with sufficient accuracy. Concerning the terms proportional to $g$, which
describe the interaction of the electromagnetic field and an ALP, they are also assumed to be small, such that only lowest-order effects in $g$ will be accounted for here.

From the above Lagrangian density the following equations of motion can be derived:
\begin{align}
\label{Eq_Phi}
 &(\square+m^2)\Phi=-\frac{g}{4\pi}(\delta_{a,s}\mathscr{F}+\delta_{a,p}\mathscr{G}),\\
\label{Eq_F}
&\partial_{\mu}F^{\mu\nu}=\frac{\alpha}{45\pi}\frac{4F^{\mu\nu}\partial_{\mu}\mathscr{F}+7\tilde{F}^{\mu\nu}\partial_{\mu}\mathscr{G}}{F_{cr}^2}
-g(\delta_{a,s}F^{\mu\nu}+\delta_{a,p}\tilde{F}^{\mu\nu})\partial_{\mu}\Phi.
\end{align}

We first consider a strong, monochromatic electromagnetic field $F_0^{\mu\nu}(x)=(\bm{E}_0(x),\bm{B}_0(x))$ with amplitude $E_0$, wave four-vector
$k_0^{\mu}=(\omega_0,\bm{k}_0)$ (angular frequency $\omega_0=|\bm{k}_0|$) and propagating along the negative $y$ direction. Thus, in this framework
$F_0^{\mu\nu}(x)=F_0^{\mu\nu}(\varphi_0)=(\bm{E}_0(\varphi_0),\bm{B}(\varphi_0))$, with $\varphi_0=(k_0x)=\omega_0(t+y)$ and
\begin{align}
\label{E_0_B_0}
\bm{E}_0(\varphi_0)=E_0\left[\hat{\bm{z}}\cos(\varphi_0)-\sigma_0\hat{\bm{x}}\sin(\varphi_0)\right],\quad
\bm{B}_0(\varphi_0)=-E_0\left[\sigma_0\hat{\bm{z}}\sin(\varphi_0)+\hat{\bm{x}}\cos(\varphi_0)\right],
\end{align}
where $\hat{\bm{x}}$ and $\hat{\bm{z}}$ are unit vectors associated with the $x$- and the $z$-axis, respectively.  Here, the parameter $\sigma_0$ can assume
the discrete values $0$ (linear polarization) and $\pm 1$ (circular polarization with  positive/negative helicity). In addition we consider a probe plane-wave
field $\mathcal{F}^{\mu\nu}(x)=(\bm{\mathcal{E}}(x),\bm{\mathcal{B}}(x))$ counterpropagating with respect to the strong field and with four-wave vector
$k^{\mu}$ (angular frequency $\omega$), i.e., $\mathcal{F}^{\mu\nu}(x)=\mathcal{F}^{\mu\nu}(\varphi)$, with $\varphi=(kx)=\omega t-|\bm{k}|y$. Note that we
implicitly assumed that the probe field is much weaker than the strong one, such that, while the strong plane wave propagates as in vacuum (refractive index
identically equal to unity), the dispersion relation for the probe field is allowed in principle to be affected by nonlinear QED and ALP effects.

\section{Birefringence and dichroism of the vacuum}

It is convenient to Fourier-transform the above eqs.~(\ref{Eq_Phi}) and (\ref{Eq_F}) and to work in the four-momenta space. By expanding the resulting
equations up to linear terms in the probe electromagnetic field, the equation for the axion field is easily solved and one obtains:
\begin{equation}
\label{phi}
\begin{split}
\phi(k)=&-\frac{g}{4\pi}\frac{iE_0}{k^2-m^2}\langle\delta_{a,s}\{(\omega+\omega_0)[\mathcal{A}_z(k+k_0)+i\sigma_0\mathcal{A}_x(k+k_0)]+(\omega-\omega_0)\\ &\times[\mathcal{A}_z(k-k_0)
-i\sigma_0\mathcal{A}_x(k-k_0)]\}-\delta_{a,p}\{(\omega+\omega_0)[\mathcal{A}_x(k+k_0)\\&-i\sigma_0\mathcal{A}_z(k+k_0)]+(\omega-\omega_0)[\mathcal{A}_x(k-k_0)+i\sigma_0\mathcal{A}_z(k-k_0)]\}\rangle.
\end{split}
\end{equation}
In this equation it is understood that the axion field $\Phi(x)$ can be represented as the sum $\Phi_0(x)+\phi(x)$, with $\Phi_0(x)$ being the solution of
the corresponding wave equation in the presence of the strong wave only. Since both the electromagnetic invariants vanish for a plane wave, we have also
assumed that $\Phi_0(x)=0$. Moreover, in eq.~(\ref{phi}) we have introduced the Fourier transform of the four-vector potential $\mathcal{A}^{\mu}(\varphi)$ of the probe field,
which in our case can be conveniently chosen as $\mathcal{A}^{\mu}(\varphi)=(0,\bm{\mathcal{A}}(\varphi))$, with $\bm{\nabla}\cdot\bm{\mathcal{A}}(\varphi)=0$.
Finally, the two poles in the function $(k^2-m^2)^{-1}$ are intended to be shifted such that the retarded ALP propagator is obtained when going back to
configuration space. It is easy to see that by transforming now the equation of the electromagnetic field, we need to compute the ALP field at $k^{\mu}\pm k_0^{\mu}$.
In general, terms like $\bm{\mathcal{A}}(k\pm 2k_0)$ will appear in the equation for $\bm{\mathcal{A}}(k)$. However, since we are interested in the propagation
of the probe electromagnetic wave for a given frequency and since the effects on the propagation of that frequency of the side-band terms shifted by $ \omega_0$
would be high-order with respect to $\alpha$ and/or $g$, we can consistently neglect these terms in our lowest-order analysis. The resulting equation for the
vector potential of the probe reads:
\begin{eqnarray}
\label{A}
&\left(n^2-1\right)\bm{\mathcal{A}}=\frac{2\alpha}{45\pi}\frac{I_0}{I_{cr}}[4(\mathcal{A}_z \hat{\bm{z}}+\sigma_0^2\mathcal{A}_x\hat{\bm{x}})+7(\mathcal{A}_x \hat{\bm{x}}
+\sigma_0^2\mathcal{A}_z\hat{\bm{z}})]\nonumber\ \\ \displaystyle&\qquad\qquad\qquad-g^2I_0\delta_{a,s}\left[\frac{\mathcal{A}_z-i\sigma_0\mathcal{A}_x}{4\omega\omega_0-m^2}(\hat{\bm{z}}+i\sigma_0\hat{\bm{x}})
-\frac{\mathcal{A}_z+i\sigma_0\mathcal{A}_x}{4\omega\omega_0+m^2}(\hat{\bm{z}}-i\sigma_0\hat{\bm{x}})\right]\nonumber\\ \displaystyle&\qquad\qquad\qquad-g^2I_0\delta_{a,p}\left[\frac{\mathcal{A}_x+i\sigma_0\mathcal{A}_z}{4\omega\omega_0-m^2}
(\hat{\bm{x}}-i\sigma_0\hat{\bm{z}})-\frac{\mathcal{A}_x-i\sigma_0\mathcal{A}_z}{4\omega\omega_0+m^2}(\hat{\bm{x}}+i\sigma_0\hat{\bm{z}})\right],
\end{eqnarray}
where $n=\vert\bm{k}\vert/\omega$ denotes the refractive index of the probe and $I_0=E_0^2/4\pi$ is the peak intensity of the strong wave. Note that, for notational
simplicity, the dependence on $k^{\mu}$ of the vector potential has been  omitted.

\subsection{The linearly polarized case}

We first consider the case of a linearly polarized strong wave ($\sigma_0=0$). In this case, we obtain from eq.~(\ref{A}) that the refractive index of the probe depends on
if it is polarized along the $z$ direction, the so-called ``parallel'' configuration, or along the $x$ direction, the so-called ``perpendicular'' configuration (note that
the strong field is polarized along the $z$ direction). The expressions of the two corresponding refractive indexes $n_{\parallel}$ and $n_{\perp}$ are
\begin{align}\label{npa}
n_{\parallel}=&1+\frac{4\alpha}{45\pi}\frac{I_0}{I_{cr}}-\delta_{a,s}\frac{g^2I_0m^2}{16\omega^2\omega_0^2-m^4},\\
n_{\perp}=&1+\frac{7\alpha}{45\pi}\frac{I_0}{I_{cr}}-\delta_{a,p}\frac{g^2I_0m^2}{16\omega^2\omega_0^2-m^4}.\label{npe}
\end{align}
The results in a constant-crossed field can be obtained starting from Eqs. (\ref{Eq_Phi})-(\ref{Eq_F}) and by setting $\omega_0=0$ in Eq. (\ref{E_0_B_0}). As expected, 
the corresponding expressions of the refractive indexes can be obtained from Eqs. (\ref{npa})-(\ref{npe}) by setting $\omega_0=0$ and by substituting $I_0\to 2I_0$. The latter 
substitution results from the fact that in the case of the oscillating wave we have neglected the side-band terms proportional to $\bm{\mathcal{A}}(k\pm 2k_0)$ and effectively 
used the average value of the intensity of the strong wave. In the limit $g=0$ the well-known results of the refractive indexes including vacuum-polarization effects elicited by the virtual electron-positron pairs are recovered
\cite{Affleck:1987gf,Dittrich}. We note that vacuum-polarization effect due to the axion field would affect only one polarization configuration, depending on if the ALP is a scalar or a
pseudo-scalar particle. The pole at
\begin{equation}
 m_0=2\left(\omega\omega_0\right)^{\nicefrac{1}{2}}\label{resonance}
\end{equation}
stems from the fact that one photon from the probe and one from the strong field can create a real axion. In fact, by indicating as $p^{\mu}$ the four-momentum of the created
axion, the mentioned divergence corresponds to the energy-momentum conservation equation $k^{\mu}+k_0^{\mu}=p^{\mu}$. Obviously, whenever the condition (\ref{resonance}) is fulfilled,
the refractive indexes diverge and our perturbative approach is not applicable. However, eqs.~(\ref{npa})-(\ref{npe}) can be still employed to explore the domain close to the
resonance in which $m=m_0\pm\Delta m$, with $\Delta m\ll m_0$, provided that the condition
\begin{equation}
\Delta m \gg \frac{1}{4}\frac{g^2 I_0}{m_0}\label{closeresonace}
\end{equation}
is fulfilled.\footnote{We note, however, that, by assuming that the next higher-order term in $g$ is of the order of the square of the lowest-order term in eqs.~(\ref{npa})-(\ref{npe}),
then the latter equations can be consistently employed only if $(g^2I_0/4m_0\Delta m)^2\ll (4\alpha/45\pi)(I_0/I_{cr})$.} By assuming to work with a high-intensity laser as those available
at the forthcoming ELI facility \cite{ELI} or at XCELS \cite{xcels} and to employ an optical probe, the expression of the resonant mass $m_0$ and the above conditions can be written in
a more transparent way as $m_0[\text{eV}]=2\sqrt{\omega[\text{eV}]\omega_0[\text{eV}]}$
and $ 3.7\,g^2[\text{GeV$^{-1}$}]I_0[10^{25}\;\text{W/cm$^2$}]/m_0^2[\text{eV}]\ll(\Delta m /m_0)\ll 1$, respectively.

We pursue our analysis by shifting the position of the  pole  in eqs.~(\ref{npa}) and (\ref{npe}) as indicated below eq.~(\ref{phi}). Thus, the two refractive indexes acquire
an imaginary part given by
\begin{equation}\label{npara}
\text{Im}(n_{\parallel})=-\delta_{a,s}\frac{\pi}{2}g^2I_0\delta(4\omega\omega_0-m^2),\qquad
\text{Im}(n_{\perp})=-\delta_{a,p}\frac{\pi}{2}g^2I_0\delta(4\omega\omega_0-m^2).
\end{equation}
In this context, the quantities $\kappa_{\parallel/\perp}\equiv 2\,\text{Im}(n_{\parallel/\perp}) \omega$ coincide with the photon attenuation coefficients in the
head-on collision of a probe photon with the strong plane wave in the corresponding mutual polarization configuration (note that here we neglect the axion-photon back conversion, which is an higher-order process in $g$). This means that the intensity $I_{\parallel/\perp}(L)$
of a probe field polarized in the parallel/perpendicular direction, will be $I_{\parallel/\perp}(L)=\exp(-\kappa_{\parallel/\perp}L)I_{\parallel/\perp}(0)$, after
propagating a distance $L$ inside the strong laser field.

It is worth noting that the photon attenuation coefficients $\kappa_{\parallel/\perp}$ are singular as  a consequence of considering a monochromatic strong field.
The divergence, however, disappears for a pulsed strong plane-wave of the form
\begin{equation}
\label{E_p}
\bm{E}_0(\varphi_0)=E_0f(\varphi_0)\cos(\varphi_0+\varphi_{\text{CEP}})\hat{\bm{z}},\quad
\bm{B}_0(\varphi_0)=-E_0f(\varphi_0)\cos(\varphi_0+\varphi_{\text{CEP}})\hat{\bm{x}},
\end{equation}
with $f(\varphi_0)$ being a (non-negative) shape function and with the CEP $\varphi_{\text{CEP}}$. In this case, instead of employing the
optical theorem, one can directly calculate the probability that a photon with four-momentum $k^{\mu}=(\omega,\bm{k})$ and polarization $\lambda=\parallel,\perp$ transforms into an ALP in passing through the strong laser field
in eq.~(\ref{E_p}). By starting from the Lagrangian density in eq.~(\ref{L}), it follows that the probability associated with the conversion process reads
\begin{eqnarray}
\mathcal{P}_{a,\lambda}=\frac{g^2}{16}\frac{I_0}{\omega_0^2}(\delta_{a,s}\delta_{\lambda,\parallel}-\delta_{a,p}\delta_{\lambda,\perp})^2\left\vert e^{i\varphi_{\text{CEP}}}G\left(\frac{m^2}{4\omega\omega_0}+1\right)+e^{-i\varphi_{\text{CEP}}}G\left(\frac{m^2}{4\omega\omega_0}-1\right)\right\vert^2 \label{P_l}
\end{eqnarray}
where $G(\varrho)=\int d\varphi \exp(i\varrho\varphi)f(\varphi)$. We mention that, in deriving this expression of the probability, one has to transform the conserving $\delta$-function $\delta(p^+-k^+-k_0^+)$, where $q^+=\varepsilon_q+q_y$ for a generic on-shell four-momentum $q^{\mu}=(\varepsilon_q,\bm{q})$, into a $\delta$-function of the form $\delta(k_y-K_{y})$, with $K_y$ resulting from the equation $p^+-k^+-k_0^+=0$, before performing the square of the $\delta$-function itself (see also \cite{Ilderton:2012qe}).

We find interesting that for a finite pulse, the probability of ALP production depends on the pulse shape and, in particular, on the CEP of the strong laser field. On the other hand, if $f(\varphi)=1$, the total probability $\mathcal{P}_{a,\lambda}$ coincides with the corresponding photon attenuation coefficient times the interaction
length $L$, once one identifies the total interaction phase $\Phi=2\omega_0 L$ in the case of a photon counterpropagating with respect to the strong laser beam. This fact provides
evidences that the side-band contributions $\bm{\mathcal{A}}(k\pm2 k_0)$ do not play a role as long as one is interested in the leading order term $\sim g^2$.

The above analysis indicates that both for a scalar or pseudoscalar ALP, the effect of the axion can in principle be isolated from that of the virtual electron-positron pairs by
measuring the intensity of the probe after the interaction. In fact, in the former (latter) case the component of the probe field parallel (perpendicular) to the electric field
of the strong wave is expected to be attenuated by a factor $\sim\exp(-\mathcal{P}_{s,\parallel}/2)$ ($\sim\exp(-\mathcal{P}_{p,\perp}/2)$) after crossing the laser pulse. Note that $\mathcal{P}_{s,\parallel}=\mathcal{P}_{p,\perp}=\mathcal{P}_l$, with
\begin{eqnarray}
&&\mathcal{P}_l=\frac{g^2}{16}\frac{I_0}{\omega_0^2}
\left\vert e^{i\varphi_{\text{CEP}}}G\left(\frac{m^2}{4\omega\omega_0}+1\right)+e^{-i\varphi_{\text{CEP}}}G\left(\frac{m^2}{4\omega\omega_0}-1\right)\right\vert^2.
\end{eqnarray}
The reduction of a component of the electric field of the probe induces a rotation
$\delta \vartheta_a$ of the probe polarization for a scalar/pseudoscalar ALP. For a probe initially polarized at an angle $\vartheta$ with respect to the $x$-axis, the rotation $\delta\vartheta_a$ of the probe polarization angle is given by
\begin{equation}
\delta \vartheta_a=\frac{1}{4}(\delta_{a,s}-\delta_{a,p})\mathcal{P}_l\sin(2\vartheta).
\end{equation}
Note that the sign of the rotation induced by the ALP depends on whether it is a scalar or a pseudo-scalar particle. Finally, in the case of a the strong Gaussian wave with
$f(\varphi)=\exp[-\varphi^2/2(\Delta\varphi)^2]$, one obtains
\begin{equation}
\label{G}
G(\varrho)=\sqrt{2\pi}\Delta\varphi e^{-\frac{1}{2}\varrho^2\Delta\varphi^2}.
\end{equation}
We want to conclude this section by pointing out  that for a spatially focused strong field, rotation of the probe polarization
are also expected from pure nonlinear QED effects \cite{DiPiazza:2006pr,BenKing:2009,BenKing:2010}.
%

\subsection{The circularly  polarized case}

The case of a circularly-polarized strong wave
\begin{align}
\label{E_circ}
\begin{split}
\bm{E}_0(\varphi_0)&=E_0f(\varphi_0)[\hat{\bm{z}}\cos(\varphi_0+\varphi_{\text{CEP}})-\sigma_0\hat{\bm{x}}\sin(\varphi_0+\varphi_{\text{CEP}})],
\end{split}\\
\label{B_circ}
\begin{split}
\bm{B}_0(\varphi_0)&=-E_0f(\varphi_0)[\sigma_0\hat{\bm{z}}\sin(\varphi_0+\varphi_{\text{CEP}})+\hat{\bm{x}}\cos(\varphi_0+\varphi_{\text{CEP}})],
\end{split}
\end{align}
is easily analyzed by considering  a circularly-polarized probe as well with helicity $\sigma$. In this case the refractive  $n_{\sigma}$ is the
same for both a scalar and a pseudoscalar ALP and, for a monochromatic strong field, it reads
\begin{equation}
n_{\sigma}=1+\frac{11\alpha}{45\pi}\frac{I_0}{I_{cr}}-\frac{g^2I_0\sigma\sigma_0}{4\omega\omega_0-\sigma\sigma_0 m^2}.\label{ncircualr}
\end{equation}
In the case where $g=0$ the resulting refractive index coincides with the one obtained in \cite{Affleck:1987gf,baier}. Besides, our expression shows that only
the refractive index of a wave with the same helicity of  the strong field ($\sigma\sigma_0=+1$) acquires an imaginary part connected to the
total production probability $\mathcal{P}_{\sigma}$ of an ALP. This fact is explained via the conservation of the  component $J_y$ of the total
angular momentum along the $y$-axis and the fact that in the monochromatic case all laser photons have helicity $\sigma_0$\footnote{This is statement is no
longer valid in the case where the external laser wave is a pulsed field of the form (\ref{E_circ})-(\ref{B_circ}).}. In fact, since the ALP also propagates along the $y$-direction and since
it is a spin-0 particle, the creation process is allowed only if the helicity of the probe and of the strong-field photon are the same. For the
same reason as for the linear polarization, this probability diverges. This fact  motivates us to investigate the realistic case where the field of the wave is
a circularly polarized laser pulse. In such a  case, the ALP production probability is finite and, both in the scalar and in the pseudoscalar case, it is given by
\begin{equation}
\label{P_sigma}
\begin{split}
\mathcal{P}_{\sigma}=&\frac{g^2}{8}\frac{I_0}{\omega_0^2}\left\vert\delta_{\sigma\sigma_0,-1} e^{i\varphi_{\text{CEP}}}G\left(\frac{m^2}{4\omega\omega_0}+1\right)+\delta_{\sigma\sigma_0,1}e^{-i\varphi_{\text{CEP}}}G\left(\frac{m^2}{4\omega\omega_0}-1\right)\right\vert^2.
\end{split}
\end{equation}
We note that the factor two arising in this expression with respect to eq.~(\ref{P_l}) does not represent an effective enhancement of the probability,
as for producing a circularly polarized wave with a given amplitude, twice the energy is required than for a linearly polarized field with the same amplitude. Also in the present circularly-polarized case, the ALP production probability in general depends on the temporal shape and, in particular, on the CEP of the strong pulse.
Due to ALP production, a linearly-polarized probe field passing through a circularly polarized strong laser beam with $\sigma_0=\pm1$ acquires an ellipticity $\psi$, given by
\begin{equation}\label{psi}
\psi=\frac{1}{4}\vert\mathcal{P}_+-\mathcal{P}_-\vert=\frac{g^2}{16}\frac{I_0}{\omega_0^2}\left\vert G^2\left(\frac{m^2}{4\omega\omega_0}-1\right)-G^2\left(\frac{m^2}{4\omega\omega_0}+1\right)\right\vert,
\end{equation}
which can be exploited as a possible observable indicating the occurrence of ALP photo-production (see also section~\ref{limits}). By employing eq.~(\ref{G}), the expression of
the ellipticity in the case of a Gaussian profile, can be easily obtained.

It is interesting to observe here that higher-order corrections to the refractive index in $\sim\omega\omega_0/m_e^2$ due to
vacuum polarization do also depend on the mutual helicity of the probe and the strong wave \cite{Mitter,mini}. A similar result has been obtained from the general
expression of the vacuum polarization tensor \cite{baier}. However, such corrections are safely negligible at the optical frequencies we have in mind here.

\section{Exclusion limits \label{limits}}

Since the birefringence effects in a strong circularly polarized wave only arise in principle from the photon-ALP interaction, we limit to this case here. 

The non-observation of astrophysical and cosmological consequences embed the mass of the QCD axion within a quite limited range $10^{-6}\;\text{eV}<m<10^{-2}\;\text{eV}$ 
\cite{Asztalos:2001jk,lamoreaux}. It is worth mentioning at this point that the lowest limit has been established from different axion production processes which 
arise in different early universe scenarios as such vacuum realignment and string decay \cite{Asztalos:2001jk}. In contrast, the  upper one follows from the absence of 
an observational discrepancy--associated with an  energy-loss argument--in the supernova SN1987a \cite{Asztalos:2001jk,lamoreaux}. In order to keep the argument as general 
as possible we employ the more general expression of $n_{\sigma}$ in eq. (\ref{ncircualr}). In this framework, the direction of polarization of an initially linearly 
polarized probe, after passing through a circularly-polarized strong laser field extending over a length $L$, will be tilted by an angle $\vartheta=(1/2)(n_+-n_-)\omega L$. 
In the case under consideration, it is
\begin{equation}
\left.\vartheta\right\vert_{m^2\neq 4\omega\omega_0}\approx -\sigma_0\frac{g^2I_0}{4\omega_0\omega-m^2}\omega L\label{ellipticity},
\end{equation}
where the condition $m^2\neq 4\omega\omega_0$ has to be intended as that we are sufficiently far away from the resonance according to eq. (\ref{closeresonace}).
Since the strong wave is assumed to be monochromatic, it is understood that $\omega_0L=2\pi L/\lambda_0\gg 1$, where we have introduced here the wavelength $\lambda_0$ 
of the strong field. Eq. (\ref{ellipticity}) indicates that the rotation of the polarization angle depends on the product $I_0\times L$. From this point of view, 
at a given strong laser spot radius, the result depends only on the total energy of the laser.

In order to investigate how the presented experimental setup can provide information on the value of the coupling constant $g$, we initially consider the parameters
of the quasi-monochromatic strong pulse as those envisaged for OMEGA EP system  \cite{OMEGA}  at Rochester, USA.  Among its  four consisting  beamlines, there are  two
which could operate in the infrared regime with  $\omega_0\approx1.17\ \rm eV$ ($\lambda_0=1\;\text{$\mu$m}$),  temporal length of about $1 \ \rm ps$ (corresponding to
$L=300\;\text{$\mu$m}$) by delivering, in addition,  intensities $I_0$ of about $2\times10^{20}\ \rm  W/cm^2$. The probe beam, on the other hand, could be one of the
remaining beamlines operating at the same frequency $\omega=\omega_0$. By taking into account that in the optical regime ellipticities of the order of $10^{-10}$
have been already measured \cite{Muroo_2003}, a negative result of such an experiment would exclude coupling constants $g\gtrsim 3.1  \times 10^{-5}\; \mathrm{GeV}^{-1}$. 
We point out that the measurement of an ellipticity of the order of $10^{-10}$ requires that at least about $10^{20}$ photons from the probe pass through the 
interaction region. This requirement is fulfilled by the OMEGA EP system, as each beamline has an energy of $1\;\text{kJ}$, i.e. about $10^{22}$ photons \cite{OMEGA}. 
In addition, since the beams have similar features and they have a relatively long duration, one can expect that, in principle, a good space-time overlapping of the 
beams can be achieved. Finally, we note that the mentioned beamlines at OMEGA EP are expected to be focused to a spot radius of $10\;\text{$\mu$m}$ \cite{OMEGA}, i.e. 
about ten times the laser wavelength, such that our calculations performed in the plane-wave approximation are expected to apply with sufficient accuracy.

Let us analyze the constraint that arises  when  an ultra-intense laser pulse of duration $15\;\text{fs}$, photon energy $\omega_0\approx1.55\;\text{eV}$ (such that $L=4.5\;\text{$\mu$m}$)
and intensity $I_0\approx10^{25}\;\text{W/cm$^2$}$ is taken into account. We point out that laser systems with such parameters are envisaged to be available at the ELI and at XCELS facilities.
In order to have a precise synchronization between the colliding  waves it will be convenient to chose the probe beam as a fraction of the high-intensity laser wave and to double its frequency 
($\omega=2\omega_0$).
Considering the sensitivity given above for polarimetric measurements in the optical regime,  we find that the values of $g\gtrsim 1.3  \times 10^{-6}\; \mathrm{GeV}^{-1}$ are excluded for 
ALP masses sufficiently far from and smaller than the resonance value $3.1 \ \rm eV$ (note that ``how far'' also depends on the value of the coupling constant $g$, see eq. (\ref{closeresonace})).
This result is more stringent than those obtained via laboratory-based  experiments, as the ``Light-Shining-Through-a-Wall'' ones \cite{Ehret:2010mh,Ehret:2009sq} for ALP masses above $10\;\text{meV}$.
It is worth observing that the energy of both the ELI and the XCELS facility will largely exceed $1\;\text{kJ}$ \cite{ELI,xcels} such that an adequate choice of the fraction of energy employed
for the probe beam will ensure in principle that more than $10^{20}$ probe photons pass through the interaction region. This would allow, at least in principle, for measuring such small ellipticities as $10^{-10}$
in a single shot, which is particularly important in this case as, according to the present knowledge, a high shot-to-shot repeatability of the laser performances at such intensities is not
guaranteed. Finally, we mention that in order to reach such high intensities, the beams at the ELI and the XCELS facility are expected to be spatially focused to about one wavelength \cite{ELI,xcels}.
Thus, we expect that our numerical results provide an order-of-magnitude estimate, which is enough here as a first investigation of such effects. However, a more realistic spatial shape of the laser
pulse has to be employed for a more quantitative prediction.

\begin{figure}
\begin{center}
\includegraphics[width=3.8 in]{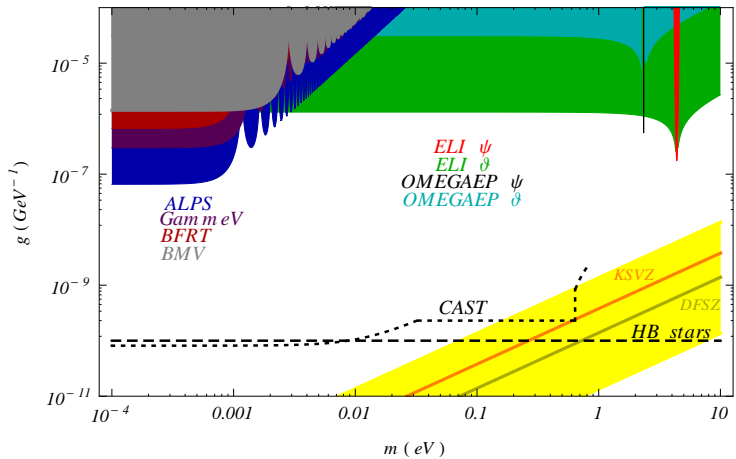}
\caption{\label{fig:mb001} Exclusion regions in the $(g,m)$-plane obtained from a polarimetric setup assisted with an intense circularly polarized laser field. 
While the green and cyan shaded areas were determined from the rotation induced by an ALP on the initial polarization plane eq.~(\ref{ellipticity}), the black and red wedges were found from 
the ellipticity eq.~(\ref{final}). The respective resonant peaks occur at $m_\mathrm{0}=2.3\ \rm eV$ (black) and $m_\mathrm{0}=4.4\ \rm eV$ (red).
The inclined yellow band covers the predictions of the axion models with $\vert E/N-1.95\vert=0.07-7$ (the notation of this formula is in accordance with Ref. \cite{CAST}). The constraint resulting from the horizontal branch stars (HB stars (dashed line)) 
are also shown. Further exclusion regions (shaded areas in the left corner) provided by  different experimental collaborations dealing with Light Shining 
Through a Wall mechanism have been included too (see legend). The exclusion limit resulting from the solar monitoring of a plausible ALP flux (CAST experiment) 
has been included as well (dotted line). We point out that the upper bound resulting from such an experiment strongly oscillates in the mass region  $0.4 \ \mathrm{eV}\leqslant m\leqslant0.6\ \rm eV$. 
Since we cannot reproduce this oscillating pattern, we show a straight dotted line, corresponding to the exclusion limit $g\leqslant2.3\times 10^{-10}\ \rm GeV^{-1}$, established 
in \cite{CAST} at $95\% $ confidence level. For the exact picture of the CAST exclusion limits, we refer the reader  to the original publication \cite{CAST} by the CAST collaboration. 
 }
 \end{center}
\end{figure}

If one ignores astrophysical ALP-mass bounds, severe constraints can be also found from the ellipticity [eq.~(\ref{psi})], when working close to the resonance for 
$\delta=|4\omega\omega_0-m^2|/m^2\ll 1$. For example, in the case of a Gaussian pulse [see eq.~(\ref{G})] with $\Delta\varphi$ larger than unity, the ellipticity in eq. (\ref{psi}) reduces to
\begin{equation}\label{final}
 \left.\psi\right\vert_{\delta\ll 1}\approx\frac{\pi^3}{2}g^2\frac{I_0}{\omega_0^2}\mathcal{N}^2e^{-(2\pi \mathcal{N}\delta)^2},
\end{equation}
where we set $\Delta\varphi=2\pi \mathcal{N}$, with $\mathcal{N}$ being the number of cycles in the pulse. By assuming to work sufficiently close to the resonance that
$2\pi \mathcal{N}\delta\ll 1$, and by considering again the parameters envisaged at ELI and XCELS (for these lasers it is $\mathcal{N}\approx 5.6$), the absence of a trace
of an ALP in an experiment with sensitivity of the order of $10^{-10}$, would imply that values $g\gtrsim 1.8\times 10^{-7}\ \rm GeV^{-1}$ are excluded 
at $m\approx 4.4\ \rm eV$. Note that eq. (\ref{final}) also suggests that more stringent exclusion limits can be obtained for very long pulses 
at relatively low intensities. However, such exclusion limits will be valid in very narrow regions according to the constraint $2\pi \mathcal{N}\delta\ll 1$. 
As an example, we show the numerical results in Fig.~\ref{fig:mb001} (black wedge) for the parameters associated with the OMEGA EP laser: $I_0= 2 \times 10^{20}\ \rm W/cm^2$, $\mathcal{N} = 300$, $\omega_0=1.17\ \rm eV$. We remark that the 
constraint $2\pi \mathcal{N}\delta\ll 1$ prevents from taking the monochromatic limit $\mathcal{N}\to \infty$ directly in eq. (\ref{final}). This limit can be obtained by 
first reconsidering eq. (\ref{psi}) and by applying the procedure described below eq. (\ref{P_l}).

Our potential discovery are summarized in Fig.~\ref{fig:mb001}, where some experimental results have been included. The prediction covered by the 
hadronic models of Dine-Fischler-Srednicki-Zhitnitskii (DFSZ) \cite{Dine:1981rt,zhitnitskii} and Kim-Shifman-Vainshtein-Zakharov (KSVZ) \cite{kim,shifman} axions are also displayed. 
Clearly, our results shows that complementary regions in the $(g,m)$-plane, which are not accessible by laboratory investigation based on dipole magnets, can be probed. 
Still, it remains excluded by the  upper bound resulting  from the solar search of axions coming from the sun \cite{CAST} and 
the constraints associated with the horizontal branch (HB) stars, emerging  as consequence of considerations of stellar energy loss due to 
the axion production \cite{Raffelt:2006cw}. However, it is worth mentioning at this point that the bounds resulting from the absence of plausible astrophysical 
and cosmological consequences rely on many assumptions which, eventually,  might over estimate the upper limit. Indeed, it has been established that the inclusion 
of macroscopic quantities such as the temperature and the density of the star might relax the aforementioned constraints \cite{evading,Gies:2007ua}.

\section{Summary and outlook}

The self interaction of the electromagnetic field, as mediated by the virtual electron-positron pairs and by a (pseudo)scalar ALP field has been analyzed. In particular, the change
in the polarization state of a probe laser field passing through a counterpropagating  strong laser field has been studied by considering the two cases of linearly and circularly polarized strong field.
In both cases a perturbative treatment was implemented to determine the refractive indexes of the probe. In the case of a linearly polarized strong beam, the coupling with the
scalar/pseudoscalar ALP leads to a modification of the already existing birefringence resulting from the polarization of the virtual electron-positron pairs. In contrast,
when the strong field is circularly polarized, the pure QED-vacuum is no longer birefringent. However, the coupling with an ALP does generate a birefringence and dichroism in the vacuum.
The latter is due to the fact that, according to the conservation of angular momentum, only a probe photon and a strong field photon with the same helicity can produce an ALP.

In addition, the photon attenuation coefficients in the case of a plane wave with arbitrary pulse form have been determined (the particular case of a Gaussian profile has been worked
out explicitly).  The resulting ellipticity induced by the photon-axion oscillation in the field of a circularly polarized plane wave has been then calculated. Both in the linearly-
and in the circularly-polarized strong-field case, we have seen that the attenuation coefficients depend on the temporal shape and, in particular, on the CEP of the strong pulse.
Moreover, we have shown that a high-precision measurement of the latter observable could improve the experimental constraints on the photon-axion coupling constant in the region in
which the ALP mass is of the order of $\sim 1 \ \rm eV$. Of course, several questions regarding the ALP-photon oscillation in a strong laser field remain open. It would be interesting,
for instance, to determine how the collision angle between the probe and the strong laser fields could be exploited to analyze the resonant masses below the eV-regime. This and other
issues will be studied in \cite{selym}.

\section*{Acknowledgments}

S. Villalba-Chavez is very grateful to Ben King, Carsten M\"{u}ller and  Hector M. Casta\~neda for helpful discussions.

\end{document}